%% file: main.tex
\documentclass[sigconf]{acmart}

\copyrightyear{2023} 
\acmYear{2023} 
\setcopyright{acmlicensed}\acmConference[Internetware 2023]{14th Asia-Pacific Symposium on Internetware}{August 4--6, 2023}{Hangzhou, China}
\acmBooktitle{14th Asia-Pacific Symposium on Internetware (Internetware 2023), August 4--6, 2023, Hangzhou, China}
\acmPrice{15.00}
\acmDOI{10.1145/3609437.3609450}
\acmISBN{979-8-4007-0894-7/23/08}

\usepackage{lineno,hyperref}
 
\usepackage{amsmath,amssymb,amsfonts}
\usepackage{float}
\usepackage{subfig}
\usepackage{graphicx}
\usepackage{url}
\usepackage{xcolor}
\usepackage{stfloats}
\usepackage{tcolorbox}
\usepackage{amsmath}
\usepackage{amsfonts}
\usepackage{multirow}
\usepackage{booktabs}
\usepackage{listings}
\usepackage{pifont}
\usepackage{makecell}
\usepackage{nicematrix}
\usepackage{bm}
\usepackage{threeparttable}
\usepackage[utf8]{inputenc}
\usepackage{ulem}

\definecolor{verylightgray}{rgb}{.97,.97,.97}

\newcommand{\tool}{APICom}
\newcommand{\atpart}{ATCom}
\newcommand{\APIcheck}{APICheck}
\newcommand{\RQone}{RQ1: How effective is our proposed approach {\tool} when compared with state-of-the-art baselines?}
\newcommand{\RQtwo}{RQ2: How effective is our proposed component {\atpart} in our proposed approach {\tool}?}
\newcommand{\RQthree}{RQ3: How does different pre-trained models affect the performance of {\tool}?}
\newcommand{\RQfour}{RQ4: Whether using prompts can help to improve the performance of {\tool}?}

\begin{document}
\begin{sloppypar}
\title{{\tool}: Automatic API Completion via Prompt Learning and  Adversarial Training-based Data Augmentation}

\author{Yafeng Gu}
\affiliation{%
\institution{Nantong University}
\country{China}
}
\email{yafeng.g@outlook.com}

 \author{Yiheng Shen}
 \affiliation{%
\institution{Nantong University}
\country{China}
}
\email{yiheng.s@outlook.com}

 \author{Xiang Chen}
 \authornote{Xiang Chen is the corresponding author.}
 \affiliation{%
\institution{Nantong University}
\country{China}
}
\email{xchencs@ntu.edu.cn}

 \author{Shaoyu Yang}
 \affiliation{%
\institution{Nantong University}
\country{China}
}
\email{shaoyuyoung@gmail.com}

 \author{Yiling Huang}
 \affiliation{%
\institution{Nantong University}
\country{China}
}
\email{hylwing13150522911@163.com}

 \author{Zhixiang Cao}
 \affiliation{%
\institution{Nantong University}
\country{China}
}
\email{486478817@qq.com}

\begin{abstract}
Based on developer needs and usage scenarios, API (Application Programming Interface) recommendation is the process of assisting developers in finding the required API
among numerous candidate APIs.
Previous studies mainly modeled API recommendation as the recommendation task, which can recommend multiple candidate APIs for the given query, and developers may not yet be able to find what they need.
Motivated by the neural machine translation research domain, we can model this problem as the generation task, which aims to directly generate the required API for the developer query.
After our preliminary investigation, we find the performance of this intuitive approach is not promising. The reason is that there exists an error when generating the prefixes of the API. However, developers may know certain API prefix information during actual development in most cases. Therefore, we model this problem as the automatic completion task and propose a novel approach {\tool} based on prompt learning, which can generate API related to the query according to the prompts (i.e., API prefix information). Moreover, the effectiveness of {\tool} highly depends on the quality of the training dataset. In this study, we further design a novel gradient-based adversarial training method {\atpart} for data augmentation,
which can improve the normalized stability when generating adversarial examples. 
To evaluate the effectiveness of {\tool}, we consider a corpus of 33k developer queries and corresponding APIs.
Compared with the state-of-the-art baselines, our experimental results show that {\tool} can outperform all baselines by at least 40.02\%, 13.20\%, and 16.31\% in terms of the performance measures EM@1, MRR, and MAP.
Finally, our ablation studies confirm the effectiveness of our component setting (such as our designed adversarial training method, our used pre-trained model, and prompt learning) in {\tool}.

\end{abstract}




\maketitle

%%%%%%%%%%%%%%%%%%%%%%%%%%%%%%%%%%%%%%%%%%%%%%%%%%%%%%%%%%%%%%%

\input{1-introduction}

\input{2-background}

\input{3-approach}

\input{4-ex_setup}

\input{5-ex_result}

\input{6-discussion}

\input{7-conclusion}

%%%%%%%%%%%%%%%%%%%%%%%%%%%%%%%%%%%%%%%%%%%%%%%%%%%%%%%%%%%%%%

\section*{Acknowledgement}
Yafeng Gu and Yiheng Shen have contributed equally to this work and they are co-first authors.
The authors would like to thank three anonymous reviewers for their insightful comments and suggestions, which can substantially improve the quality of this work. 
This work is supported in part by the Innovation Training Program for College Students (2023214, 2023356).

\normalem

\bibliographystyle{ACM-Reference-Format}
\bibliography{main}

\end{sloppypar}
\end{document}

%% file: 1-introduction.tex
\section{Introduction}
\label{sec:intro}

API can provide developers with the ability to utilize pre-built functions and integrate disparate software systems, which can improve the efficiency and effectiveness of software development and maintenance.
Therefore, API (Application Programming Interface) has become an important component of modern software development. 
However, with the rapid growth of available APIs and the increasing complexity of software systems, it has become difficult for developers to find the most suitable APIs based on their needs.
Therefore, effective API recommendation approaches can help to improve the efficiency and effectiveness of software development and maintenance.

According to our statistical analysis, 10,870 posts can be searched by using the keyword "API" on Stack Overflow until March 2023.
In a post\footnote{\url{https://stackoverflow.com/questions/5671386/pdf417-image-generation-api-recommendation}} shown in 
\tablename~\ref{tb:table1}, the developer wants to find a corresponding API in the Grails plug-in or Java library, which can generate PDF417 images in the Grails application and send them by email. This post receives more than 12.8k likes, and the first answer about API receives more than 75.2k  likes. Based on the above analysis, we can find it challenging for developers who are unfamiliar with the development task at hand to find the required API quickly.

\begin{table}[h]
\centering
\setlength{\tabcolsep}{5mm}
\scriptsize
	\caption{A post related to API recommendation from Stack Overflow}
	% \vspace{-0.3cm}
	\begin{tabular}{cl}
	\toprule
    \textbf{Post Title}    &  PDF417 image generation API recommendation     \\ \midrule
	\multirow{4}*{\textbf{Post Content}}	& in a grails application I need to generate a PDF417 image\\
 &    and send it via email. Can   anybody recommend me \\
 &  a (hopefully free)  Grails plugin or Java library? \\ \midrule
 % &   \\ \midrule
    \textbf{Post Tags}  & \ java , \  grails  , \ barcode        \\ \bottomrule
\end{tabular}  
% \vspace{-0.35cm}
\label{tb:table1}
\end{table}

\begin{figure}[htbp]
	\centering
    % \vspace{-0.25cm}
	\includegraphics[width=\linewidth]{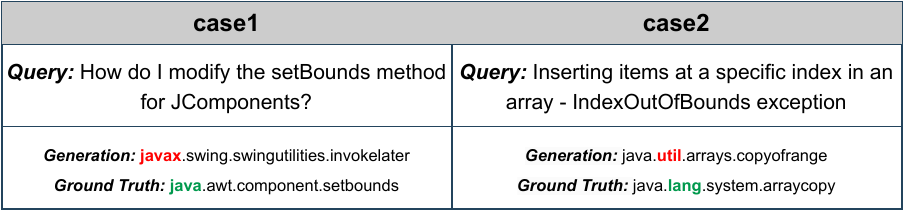}
 % \vspace{-0.3cm}
	  \caption{Two examples to show the limitations of using the intuitive generation approach for API recommendation}
    % \vspace{-0.25cm}
	\label{fig:case1}
\end{figure}

To solve this problem, automatic API recommendation approaches have been studied. Most of the previous studies mainly modeled API recommendation as the
information retrieval (IR) task, which can recommend multiple candidate APIs based on text similarity.
However, the developers still need to select the desired API from these candidate APIs, and sometimes even the correct API is not included in these candidate APIs.
Motivated by the neural machine translation research domain, one intuitive way is to model this problem as the automatic generation task, which can generate the required API for the developer's query directly.
Although the popular generation models perform well in software engineering generation tasks (such as source code summarization~\cite{ahmad2020transformer,li2022setransformer,yang2022ccgir,li2021secnn,yu2022bashexplainer}, issue title generation~\cite{chen2020stay,lin2023gen}, code generation~\cite{yang2022dualsc,yang2023exploitgen}, Stack Overflow title generation~\cite{liu2022sotitle,zhouqtc4so}), we find that the performance of this intuitive way is not promising after our preliminary investigation.
% \gyf{Specifically, we can use sequence-to-sequence models to learn the mapping between the developer query and the relevant API. This approach has several advantages over traditional recommendation methods, such as the ability to generate new APIs that do not exist in the training data and the ability to handle out-of-vocabulary terms. However, this approach also presents challenges, such as the need for large amounts of high-quality training data and the risk of generating incorrect or irrelevant APIs. To overcome these challenges, we propose a novel approach that combines the strengths of both the IR and generation methods, leading to more accurate and effective APIs for developers.}
% After a preliminary investigation, we find the performance is not promising.
% Unlike natural language descriptions, the content of the API is uniform and the API should be fully correct to be used by developers.
% 引出生成例子
\figurename~\ref{fig:case1} shows two examples of generating incorrect APIs by using this intuitive generation approach. 
% Based on these two examples, we can find that the content between the two APIs is similar, but the generated API is incorrectly formatted and unusable. Therefore, generating the correct API that can be used is the primary challenge.
In the first example, the ground truth is "java.awt.component.setbounds". If we use the intuitive generation approach, the generated API is "java.swing.swingutilities.invokelater". In this case, there is an error with the first prefix.
In the second example, the ground truth is "java.lang.system.arraycopy". If we use the intuitive generation approach, the generated API is "java.util.arrays.copyofrange". In this case, there is an error with the second prefix.
Based on these two examples, we find that incorrectly generated prefixes may cause the low performance of this intuitive generation approach.
However, based on our observation of practical software development and maintenance, developers may remember certain API prefix information and it is the remaining parts of the API that they really need to turn to for help. 
% For example, it is obvious in the Java API that the first prefix of the API is often "java" or "javax". In example 2, developers can often know that the correct API interface belongs to `java.lang', because this is the core class library in the Java language and is related to array operation.

% 推荐、生成都不行，但是补全可以的例子

% 生成的例子写完了，引出补全的例子
Based on the above motivation, we are the first to model the API recommendation problem into the API completion task and then propose a novel approach {\tool}. In particular, developers only need to input the description of the encountered problem (i.e., query) and certain API prefixes they know. After that, the corresponding correct API may be generated. To better complete the API, we consider prompt learning, which can predict the content of MASK (i.e., the remaining parts of the API) more accurately according to limited prompts.
Specifically, prompt learning is a technique that provides a small amount of context to a language model to guide its predictions.
In our API completion task, prompt learning can improve performance and prevent catastrophic forgetting through generated task-specific prompts.

On the other hand, the performance of {\tool} highly depends on the quality of the corpus used for training. 
However, collecting such data manually can be time-consuming and labor-intensive.
To alleviate this problem, we propose a novel adversarial training method {\atpart}, which can improve the stability of
the normalization process while retaining the advantages of $L_1$ and
$L_2$ normalization. 
% {\atpart} adds disturbance to the input to generate $K$ adversarial examples.
% Then these adversarial examples and the original inputs can be used to fine-tune the pre-trained model CodeT5~\cite{wang2021codet5}.
% To our best knowledge, adversarial training for data augmentation has not been thoroughly studied in the API recommendation task. To fill this gap, we introduced adversarial training innovatively.

% In this study, we are the first to model this problem as the completion problem and propose a novel approach {\tool}, which can automatically provide the completion API by giving the description of the API and some prefixes of the API. 
To evaluate the effectiveness of {\tool}, We consider a high-quality corpus shared by Huang et al.~\cite{huang2018api}, which contains API queries and corresponding APIs. 
% And each selected data will automatically generate variants according to different rules to generate different prompts.
To generate the prompts, we randomly mask a certain number of words (separated by dots) at the end of the APIs.
Then, {\atpart} generates $K$ adversarial examples for all the inputs to augment the training set.
Finally, we fine-tune a pre-trained model CodeT5~\cite{wang2021codet5} to learn the potential API completion patterns.
% To evaluate the effectiveness of our proposed approach, we conduct experiments on a corpus shared by Cai et al.~\cite{cai2019biker}.
% To verify the effectiveness of {\tool}, we consider two performance measures (i,e., MRR~\cite{radev2002evaluating} and MAR~\cite{schutze2008introduction}), which have been widely used in previous API recommendation studies~\cite{huang2018api,rahman2016rack,wei2022clear}.
% However, due to the short and clear characteristics of the API, these measures widely used for information retrieval~\cite{huang2018api,rahman2016rack,wei2022clear,lam2015combining,zanjani2015automatically} are difficult to accurately evaluate our API completion task by generation technology. Therefore, we consider using the EM metric~\cite{kenton2019bert,clark2020electra}, which measures whether the generated API by a model exactly matches the ground-truth API.
% To evaluate the effectiveness of our proposed approach, we conduct experiments on the corpus shared by Huang et al.~\cite{huang2018api}, which contains a large number of \gyf{user queries and corresponding APIs} for evaluating API recommendation tasks.
% Furthermore, we also select some suitable evaluation measures (i.e., MRR, MAP, and EM). 
Since most of the previous studies~\cite{huang2018api,rahman2016rack,wei2022clear,lam2015combining,zanjani2015automatically} mainly modeled API recommendation as the recommendation task, we first consider MRR (Mean Reciprocal Rank)~\cite{radev2002evaluating} and MAP (Mean Average Precision)~\cite{schutze2008introduction} as model performance evaluation measures.
% These metrics have been widely used in previous API recommendation research~\cite{huang2018api,rahman2016rack,wei2022clear}, therefore we select them as the primary evaluation measures.
Moreover, since we model this problem as an automatic completion task in this study. we second consider EM (Exact Match)~\cite{kenton2019bert,clark2020electra} measure, which measures whether the completed APIs exactly match the ground-truth APIs.
% \gyf{We did not choose to use the BLEU or ROUGE metrics because they are typically used to measure the similarity between two texts, which is not as relevant in the context of API recommendation. In API recommendation, it is more important to ensure that the recommended API accurately meets the developer's needs, rather than to measure the similarity between the recommended API and the ground-truth API. Therefore, we believe that the EM metric is a more appropriate choice for evaluating the performance of our API recommendation system.}

In our study, we want to answer the following four research questions (RQs).

\textbf{{\RQone}}

\textbf{Results.}
% Since there are no previous studies on API completion, we consider API recommendation approachs (i,e., BIKER~\cite{huang2018api}, RACK~\cite{rahman2016rack}, and CLEAR~\cite{wei2022clear}) and pre-trained language models (i,e., CodeBERT~\cite{feng2020codebert}, UniXcoder~\cite{guo2022unixcoder} and PLBART~\cite{ahmad2021unified}) as baselines. The final results show that {\tool} outperforms these two types of baselines in recommending the APIs that users need.
In our study, we first consider the classical API recommendation approaches (i,e., BIKER~\cite{huang2018api}, RACK~\cite{rahman2016rack}, and CLEAR~\cite{wei2022clear}) as the baselines. Moreover, since we are the first to model this problem as the automatic completion task, we also consider pre-trained models (i,e., CodeBERT~\cite{feng2020codebert}, UniXcoder~\cite{guo2022unixcoder} and PLBART~\cite{ahmad2021unified}) as the baselines. Our experimental results show that {\tool} outperforms both types of baselines in recommending the APIs that developers require.

\textbf{{\RQtwo}}

\textbf{Results.}
To show the effectiveness of {\atpart}, we consider three different classical adversarial training methods for data augmentation (i.e., FGSM~\cite{goodfellow2015explaining}, FGM~\cite{miyato2017adversarial}, PGD~\cite{madry2017towards}).
Based on our ablation results, we find using {\atpart} can help to achieve the best performance for {\tool}.

\textbf{{\RQthree}}

\textbf{Results.} 
To show the effectiveness of using CodeT5 in {\tool}, we consider three different pre-trained models (i.e., PLBART~\cite{ahmad2021unified}, UniXcoder~\cite{guo2022unixcoder}, CodeBERT~\cite{feng2020codebert}). 
Based on our ablation results, we find using CodeT5~\cite{wang2021codet5} can help to achieve the best performance for {\tool}.

\textbf{{\RQfour}}

\textbf{Results.} 
% When conducting natural language processing tasks, using prompts can help models better understand input text and improve model performance. However, the length of prompts may affect model performance. Therefore, it is necessary to investigate the impact of prompt length on the performance of {\tool}.
% we investigate the performance impact of prompts with different lengths (such as no prefix, one prefix, two prefixes, or more prefixes). 
% % The final results demonstrate that prompts with random words can achieve the best performance.
% The final results demonstrate that prompts with random words may outperform traditional prompts.
We conduct research to investigate the impact of using prompts on the performance of {\tool}, and our results show that {\tool} with prefixes can help to achieve the best performance.

Compare to previous studies on API recommendation~\cite{huang2018api,rahman2016rack,wei2022clear}, our study investigates a more practical problem. By providing some prompts with the query, our proposed approach {\tool} can help to provide accurate API completion suggestions, which can eventually generate APIs developers require.

To our best knowledge, the main contributions of our study can be summarized as follows.

\begin{itemize}
    \item \textbf{Direction.} 
    % We conducted novel research on the API completion task based on natural language queries from users and utilization of a subset of the API library, which opens up a new direction for API recommendation.
    We are the first to study the task of API completion, which opens a new direction for automatic API recommendation.
    
    \item \textbf{Approach.} To solve this task, we propose a novel approach {\tool} based on  natural language queries from developers and prefixes of APIs. {\tool} adopts the prompt learning paradigm and adversarial training-based data augmentation, and leverages the pre-trained CodeT5 model for learning API completion patterns automatically.
    
    \item \textbf{Evaluation.} 
    % \cx{show experimental subjects. and main findings} 
    % 和baseline比较：传统API推荐方法及预训练模型方法
    We use a corpus containing 33k API queries and corresponding APIs as our experimental subjects. The extensive and comprehensive empirical study shows the competitiveness of our proposed approach {\tool} and the component setting rationality (such as the prompts, the adversarial training method, and the pre-trained model) in {\tool}.

    %\item \textbf{Dataset.} We have publicly released the source code of the PL2AC model and the experimental dataset we used, with the aim of assisting other researchers in replicating and extending our research.
    % 我们已经公开发布了PL2AC模型的源代码和我们所用的实验数据集，旨在协助其他研究人员对我们的研究进行复现和拓展。

\end{itemize}

To promote the replication of our research and encourage more follow-up studies in this direction, we share our corpus and scripts on our project homepage\footnote{\url{https://anonymous.4open.science/r/APICom}}.

The rest of this paper is organized as follows.
Section~\ref{sec:bg} provides research background (such as API recommendation and data augmentation technology) and illustrates our research motivation. 
Section~\ref{sec:approach} describes the framework and details of our proposed approach {\tool}. 
Section~\ref{sec:setup} shows empirical settings. 
Section~\ref{sec:result} presents our analysis for different research questions. 
Section~\ref{sec:discussion} discusses related studies to our work and shows the novelty of our study. 
Finally, Section~\ref{sec:conclusion} summarizes our work and shows potential future directions.

%% file: 2-background.tex
\section{Background and Research Motivation}
\label{sec:bg}

In this section, we first introduce the background to the task of API recommendation and data augmentation. 
After analyzing these related studies, we emphasize the research motivation of our study.

\subsection{API Recommendation}

API recommendation task refers to recommending APIs suitable for developers based on their needs and historical behavior. 
The main purpose of the API recommendation task is to help developers use and integrate APIs more efficiently and improve the efficiency of their software development and maintenance.
% The previous API recommendation methods can be divided into two categories: information retrieval-based methods and deep learning-based methods.
Gu et al.~\cite{gu2016deep} proposed DeepAPI, which can generate API usage sequences for a given query.
Rahman et al.~\cite{rahman2016rack} proposed RACK, which recommends relevant APIs for a query by exploiting keyword-API associations from the crowdsourced knowledge of Stack Overflow.
Huang et al.~\cite{huang2018api} proposed BIKER. Specifically, BIKER first uses the word embedding technique to calculate the similarity between two queries. Then BIKER leverages Stack Overflow to extract candidate APIs, and ranks candidate APIs by considering the query’s similarity with the information from both Stack Overflow posts and API documentation. 
Wei et al.~\cite{wei2022clear} proposed CLEAR.
this approach first embeds the whole sentence of queries and
Stack Overflow  posts with a BERT-based model. Then CLEAR uses
contrastive learning for
learning precise semantic representations of programming terminologies.
Finally, CLEAR trains a re-ranking model to optimize its recommendation results
% Hadi et al.~\cite{hadi2022effectiveness} discussed the possibility of pre-trained models (PTMs) for API recommendation through a dataset collected from GitHub to Empirically evaluate PTMs.

% For the former, Rahman et al.~\cite{rahman2016rack} proposed an IR-based (information retrieval) method RACK, which used keywords and API associations in stack overflow crowd-sourcing knowledge to recommend a list of related APIs for natural language queries.
% Then, Huang et al.~\cite{huang2018api} proposed a method BIKER, which used word embedding technology to bridge the vocabulary gap and used API to supplement the information to bridge the knowledge gap.
% Recently, Wei et al.~\cite{huang2018api} proposed a method CLEAR, which used a BERT-based model to embed whole sentences of queries and stack overflow (SO) posts and uses contrastive learning to train the model.

% For the deep learning-based methods,
% DeepAPI~\cite{gu2016deep} is the first to introduce deep learning technology into API recommendation, which encodes a word sequence (user query) into a fixed-length context vector and generates an API sequence based on the context vector.
%ICPC那篇研究说下。
% Recently, Hadi et al.~\cite{hadi2022effectiveness} discussed the possibility of pre-trained models (PTMs) for API recommendation through a dataset collected from GitHub to Empirically evaluate PTMs.

\subsection{Data Augmentation}

Data augmentation has been widely used in the NLP field, including rule-based methods and gradient-based methods.

For rule-based methods, researchers augmented the training data by using heuristic rules.
For example, Wei et al.~\cite{wei2019eda} proposed the EDA method, which consists of four simple but effective operations (such as synonym substitution, random insertion, random exchange, and random deletion).
Xie et al.~\cite{xie2020unsupervised} proposed an unsupervised data augmentation method UDA. This method uses non-core word replacement technology, which replaces a certain proportion of unimportant words in the text with unimportant words in the dictionary.

% Recently, Claude Coulombe et al.~\cite{coulombe2018text} proposed the EDA method, which is a text surface transformation technology.
% He used simple pattern matching applied by regular expressions to provide a technique to change speech form from contraction to expansion.
For gradient-based methods (e.g., adversarial training),
researchers augmented the training data by  generating adversarial examples.
The general principle of adversarial training can be summarized as the following maximum-minimum formula:

\begin{equation} \label{eq:MaxMin}
   \underset{\theta}{min}\mathbb{E}_{(x,y)}\sim D\left [\underset{\left \| \delta  \right \| \le \epsilon }{max}  L(f_{\theta}(X+\delta),y) \right ]
\end{equation}
where $x$ represents input, $\delta$ represents perturbation, $y$ represents the label of the example, $max(L)$ represents the optimization objective, $D$ is the training set, and $\mathbb{E}$ is the maximum likelihood estimation.
% Miyato et al.~\cite{miyato2017adversarial} first introduced adversarial training and virtual adversarial training \cite{miyato2018virtual} to improve the performance of the classification model.
% Zhu et al.~\cite{zhu2020freelb} proposed FreeLB for the language model, which promotes higher invariance in embedding space by adding hostile disturbance to word embedding and minimizing input samples.
% % \gyf{Ian et al. ~\cite{goodfellow2015explaining} outputs a high-confidence incorrect answer when small but intentionally worst-case perturbations are applied to examples in the dataset.
% % Madry et al.~\cite{madry2017towards} study the adversarial robustness of neural networks from the perspective of robust optimization.
% % Miyato et.al. ~\cite{miyato2017adversarial} extend adversarial and virtual adversarial training to the text domain by applying perturbations to word embeddings in recurrent neural networks, rather than the original input itself.}
% Recently, Zhou et al.~\cite{zhou2021adversarial} trained two separate encoder-decoder models for source code sequence and abstract syntax tree (ASTs) and achieved the best performance by using antagonistic training and ensemble learning techniques.
For this kind of method, researchers mainly utilize  classical adversarial training methods (such as FGSM~\cite{goodfellow2015explaining}, FGM~\cite{miyato2017adversarial}, PGD~\cite{madry2017towards})  for data augmentation, and the effectiveness of adversarial training in the NLP field has been widely verified~\cite{morris2020textattack,yoo2021towards,li2018generating,li2021token,mao2022enhance}.

\subsection{Research Motivation}

\begin{figure}[htbp]
	\centering
    % \vspace{-1mm}
	\includegraphics[width=\linewidth]{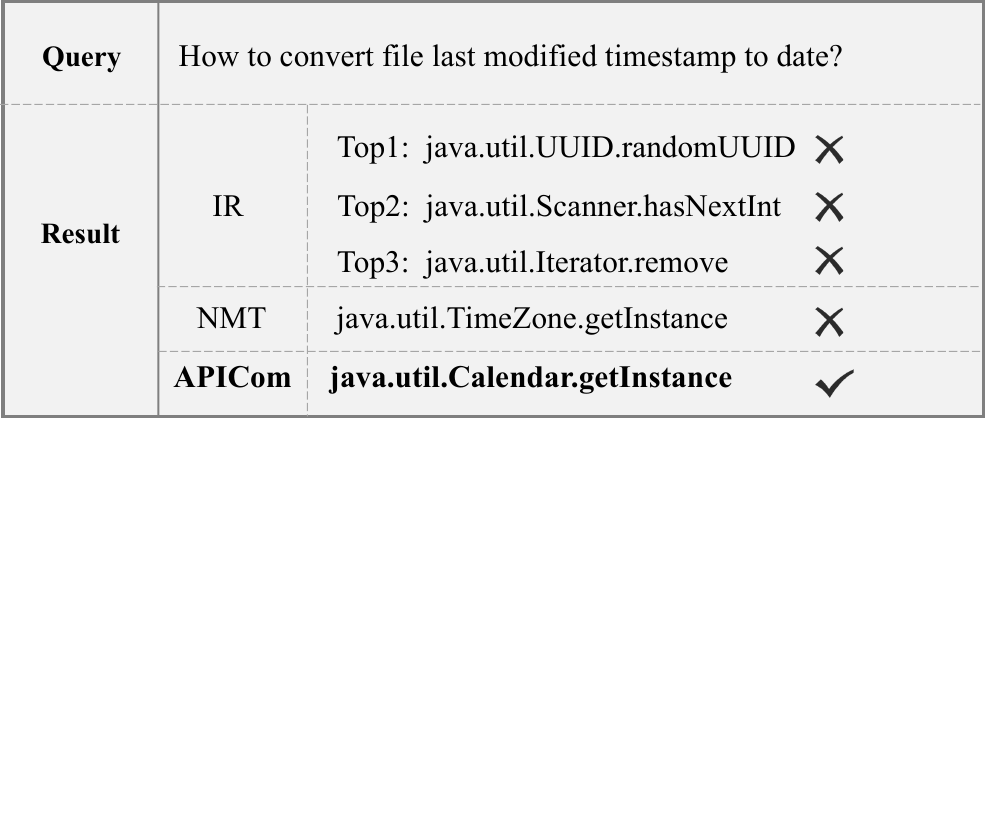}
     \vspace{-3.9cm}
        \caption{An example to show the motivation of our study}
     % \vspace{-0.5cm}

	\label{fig:case2}
\end{figure}

In this subsection, we use an example in \figurename~\ref{fig:case2} to show the motivation of our study.
In this figure, the query is "How to convert file last modified timestamp to date?".
If we use the IR-based approach BIKER~\cite{huang2018api}, the recommended top-3 APIs are "java.util.TimeZone.getInstance", 
"java.util.Scanner.hasNextInt", and "java.util.Iterator.remove".
After a manual examination, we find the required API does not exist in these top-3 APIs.
If we model this task as the automatic generation problem motivated by neural machine translation (NMT) and use the pre-trained model CodeBERT~\cite{feng2020codebert}, the generated API is "javac.util.TimeZone.getInstance". Similar to the examples in \figurename~\ref{fig:case1}, there is an error with the third prefix.
However, if the developer can provide the prefix of the required API (i.e., java.util.Calendar), our proposed approach {\tool} can generate the correct API. Based on this motivation example, we can find that API completion has great potential for effective API generation.
Therefore, combined with the prompt learning paradigm, we propose an API completion approach {\tool}.
Moreover, the quality of the dataset is vital for the effectiveness of {\tool}. In this study, we resort to the adversarial training method for performing data augmentation and design an effective method {\atpart}.  {\atpart} can augment the training set by generating $K$ adversarial examples for each original input.
The novelty of {\atpart} is that this method can improve the stability of the normalization process while retaining the advantages of $L_1$ and $L_2$ normalization.

%% file: 3-approach.tex
\section{Our Proposed Approach}
\label{sec:approach}

\begin{figure*}[htbp]
   \centering
	\includegraphics[scale=0.825]{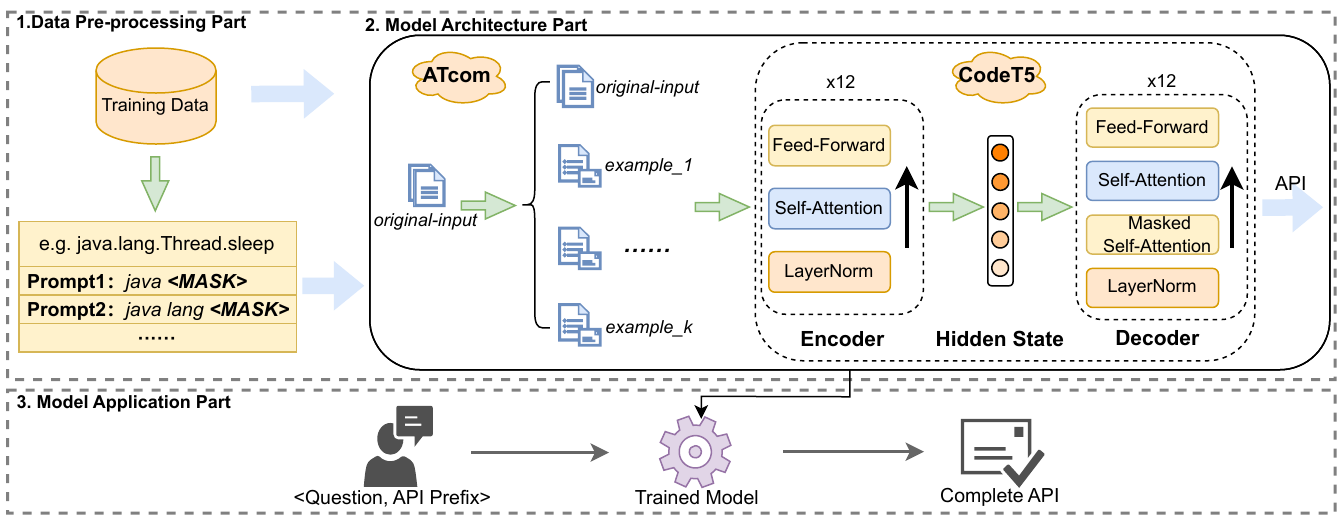}
 % \vspace{-0.4cm}
	\caption{Framework of our proposed approach {\tool}}
	\label{fig:approach} 
 % \vspace{-0.2cm}
\end{figure*}

We show the framework of our proposed approach {\tool} in \figurename~\ref{fig:approach}.
Our approach consists of three parts: data pre-processing part, model architecture part, and model application part. In the rest of this section, we present detailed information for each part.

\subsection{Data Pre-processing Part}
% In this part, we first use random MASK to construct prompt variants. and then use adversarial training to generate adversarial examples to expand the training set.

To better simulate the scenario of API completion, we  first use random masking to construct different prompts. Then we add a $<mask>$ tag after the incomplete API to indicate the masked part of the API, which can be predicted by the model trained by {\tool}.

% To better simulate the scenario in that developers enter the first half of the API in the software development process, the system automatically completes the API. 
% In this part, to improve the efficiency and accuracy of the algorithm. we first use random masking to construct different prompts. Then we add a $[mask]$ tag after the incomplete API to indicate the masked part of the API, which can be predicted by our model trained by {\tool}.

% \subsubsection{Prompt Learning}

Prompt learning~\cite{shin2020autoprompt,li2021prefix} can make pre-trained models bring the downstream task closer to the upstream task according to the specified prompt templates.
According to the position of slots, templates can be divided into $Prefix$ mode and $Cloze$ mode. The former is that the unknown slot is at the beginning of the template, and the latter is that the unknown slot is at the uncertain position of the template.
Researchers may construct different templates for the same task, and the choice of template plays an important role in the task of the prompt. Even the difference in a word can lead to a difference of more than 10 points in performance~\cite{liu2021gpt}.
In our study, we generate different prompts through $Prefix$ mode.

% \subsubsection{The Prompt Construction}

To better train the API completion model, we need to construct different prompts.
Assuming that the API $t_i$ contains $n$ words (separated by dots), we can define the masking operator as follows. Specifically, this operator first generates a random number $n_{rand}$ (1$<$$n_{rand}$$<$$n$). Then the operator generates an incomplete API (i.e., a prefix prompt) with the last $n_{rand}$ words masked.
Examples of generating incomplete APIs are shown in the data pre-processing part of Figure~\ref{fig:approach}.

% We illustrate the rationality of our designed masking operator as follows. 
% First, since developers always compose APIs in a linear way (i.e., from the first word to the last word), the masking operator selects the last consecutive words to mask.
% Second, since our study mainly concerns the API completion task, the incomplete APIs should at least contain one word. Therefore, the masking operator can  at most  mask $n-1$ words.
% In our study, we generate three different incomplete titles for each API by applying the masking operator with three different values of $n_{rand}$, and the reasons are illustrated as follows.
% First,  using Rule 4 can ensure that each API can generate at least three different incomplete APIs.
% Second, generating more incomplete APIs may help to improve model performance, but also significantly increase the model construction cost.

The rationale for our designed masking operator can be summarized as follows. Firstly, since developers typically compose APIs in a linear manner, starting from the first word and moving toward the last word, our masking operator selects the last consecutive words to be masked. Secondly, our study mainly focuses on the API completion task, which requires incomplete APIs to have at least one prefix. Therefore, the masking operator can mask at most $n-1$ words. To generate incomplete APIs, we apply the masking operator with three different values of $n_{rand}$, which can result in three different incomplete APIs for each API. Although generating more incomplete APIs could potentially improve the performance of the trained model, it would also substantially increase the training data size and eventually increase the model construction cost.

Then, we take the generated prompt and the original query as the input for our proposed model. Therefore, the considered input can be expressed as $X=prompt\oplus X_{desc}$.

\subsection{Model Architecture Part}

In this part, {\atpart} automatically generates $k$ adversarial examples according to the model input in the first phase. Then the generated adversarial examples and the original inputs are used as the training data to fine-tune our proposed API completion model in the second phase.

\subsubsection{Data Augmentation Phase}

Data augmentation can generate more training data by performing a series of transformations and extensions on the original corpus.
Specifically,
our proposed novel {\atpart} can generate $k$ adversarial examples by adding perturbation to the embedding matrix for data augmentation.

% \gyf{Data augmentation is a common technology in the field of natural language processing (NLP). 
% Its main purpose is to generate more and richer training data by performing a series of transformations and extensions on the original corpus, thereby improving the performance of the model. 
% The main goal of this method is to generate more high-quality training samples to improve the accuracy and robustness of the model and reduce the overfitting problem.}
% % Adversarial training can improve model performance and generalization in data augmentation for API completion research. Combining it with prompt learning is an effective method for further improving model performance.
% \gyf{Here, we introduce two methods: Word embedding and ATCom.}

\textbf{Word Embedding.}
Word embedding can be used to capture the relationship between tokens by mapping text to the vector representation. For the given input $X$, {\tool} tokenizes this input by the BPE algorithm~\cite{xu2022feature} to obtain the sequence $x = (x_{1}, \cdots, x_{N})$, where $N$ is the length of the sequence.
To unify the length of the input sequence, we use the fill or truncation operation. In particular, we assume the maximum input length is $n$. Then if the input sequence size is less than $n$, we complete it with 0. Otherwise, if the input sequence size is more than $n$, we truncate the excess parts. 
After the above process, we can guarantee that the output of the word embedding layer is consistent $\boldsymbol{x} = (\boldsymbol{x_{1}}, \cdots, \boldsymbol{x_{n}})$.

\textbf{{\atpart}.}
Like most adversarial training methods used in the NLP (natural language processing) domain~\cite{goodfellow2015explaining,miyato2017adversarial,madry2017towards,zhu2020freelb,jiang2020smart}, we also add perturbations to the embedding layer.
Specifically, for a given input embedding sequence $\boldsymbol{x}$, {\atpart} aims to add perturbation $\delta$ and generate $K$ adversarial examples $\left\{\boldsymbol{x_{adv}}^i\right\}_{i=1}^K$. Then CodeT5 utilizes the augmented training set (i.e., original input $\boldsymbol{x}$ and adversarial examples $\left\{\boldsymbol{x_{adv}}^i\right\}_{i=1}^K$) to fine-tune the model. 

Like Formula~(\ref{eq:MaxMin}), the purpose of our designed method {\atpart} is to find the most suitable perturbation.
Different from previous studies~\cite{goodfellow2015explaining,miyato2017adversarial}, {\atpart} does not directly calculate perturbations based on the parameter $\epsilon$, but iterates multiple times to find the optimal perturbation.
The input gradient $g_t$ of each step is calculated as follows:

\begin{equation}
g_t=\nabla _{x_t} L(f_\theta (x_t),y)
\end{equation}
where $\theta$ denotes the model parameter value, $x$ denotes the model input, $y$ denotes the label, and $L()$ denotes the loss function for model training.
Then the perturbation of each step $\delta _{t+1}$ can be computed as follows:

\begin{equation}
\delta _{t+1}=\epsilon (g_t \cdot \left\|g_t\right\| /|g_t|)
\end{equation}
where $\epsilon$ is the constraint of perturbation (i.e., $L2$ norm of the distance between the original example and the adversarial example is always $\epsilon$).
Finally, {\atpart} takes the average gradient $g_{avg}$ of $K$ times of iteration when updating parameters' values, where
\begin{equation}
g_{avg}=(g_1 + \cdots +g_t)/t
\end{equation}

Simon et al.~\cite{simon2019first} found that the adversarial loss can be expressed as a special regularization term as follows:

\begin{equation}
\widetilde{L} (x,y) \approx  L (x,y) +\frac{\epsilon }{2} \left \| \partial _xL \right \| _q
\end{equation}
where $\widetilde{L}$ represents the confrontation loss, $L$ represents the original loss of the model, and $\frac{\epsilon }{2} \left \| \partial _xL \right \| _q$ represents the special regularization term. Based on their study, we aim to use $L1$ and $L2$ regularization methods to optimize the adversarial loss.
The advantage of {\atpart} is that we use $L_1$ normalization to reduce the effect of large values on the vectors, and then apply $L_2$ normalization to ensure that the resulted vectors have a consistent length and can be summed to 1.
Therefore, {\atpart} can improve the stability of the normalization process while retaining the advantages of $L_1$ and $L_2$ normalization.

\subsubsection{Model Fine-tuning Phase with CodeT5}

CodeT5~\cite{wang2021codet5} is a unified pre-trained encoder-decoder Transformer model that better leverages the code semantics conveyed from the developer-assigned identifiers.
In this work, we adapt CodeT5 as our backbone model to solve the API completion task.

\textbf{Encoder.}
The encoder module includes 12 sub-blocks and each block is made up of two subcomponents (i.e., a self-attention layer and a small feed-forward network).
Specifically,
Self-attention~\cite{cheng2016long} is calculated by using queries ($Q$), keys ($K$) with the dimension $d_{k}$, and values ($V$) with the dimension $d_{v}$. In particular, the dot product of the queries and keys is first computed. Then the weight of each value is then calculated using the softmax function after each has been divided by $\sqrt{d_{k}}$. The output matrix can be calculated as follows:

\begin{equation} \label{eq:attention1}
   Attention(Q,K,V)=softmax(\frac{QK^{T}}{\sqrt{d_{k}}})V
\end{equation}

The feed-forward network (FFN) is composed of two linear transformations, which are separated by a ReLU activation and can be computed as follows:

\begin{equation} 
\label{eq:attention2}
   FFN(x)=\max (0,xW_{1}+b_{1})W_{2}+b_{2}
\end{equation}

Layer normalization~\cite{ba2016layer} is applied to the input of each subcomponent.
Following the layer normalization, a residual skip connection~\cite{he2016deep} adds each subcomponent’s input to its output.
The output of the encoder is defined as a hidden state vector $\boldsymbol{X_\text{out}}$.

\textbf{Decoder.}
The decoder is similar to the encoder in structure except that it includes a masked self-attention layer, the purpose of which is to prevent the model from noticing unknown information during model training.
In the decoding step, the model can generate the following API tokens by the encoder output $\boldsymbol{X_\text{out}}$.
%Suppose in the ($k$+$1$)-$th$ decoding step, we can generate the first $k$ tokens of the predicted API\cx{?} and then input the embedding vector of the generated sequence and the encoder output $\boldsymbol{X_\text{out}}$ to the decoder, the model will generate the ($k$+$1$)-$th$ token.
The network can predict the probability of the next token through the softmax layer, which can be defined as follows.

\begin{equation} \label{eq:decoder1}
   P(y_{t+1} \mid y_{1}, \ldots , y_{t}) = softmax(W \cdot X_\text{out} + b)
\end{equation}
where $y$ denotes the predicted token. We train our model parameters $\theta$ by the loss function $L$ for a given input text $x$ based on cross-entropy, which is defined as follows.

\begin{equation} \label{eq:decoder2}
   L = -\sum_{i=1}^{\left | y \right |} log P_{\theta}(y_{i} \mid y<i, x)
\end{equation}

Since the output of our proposed approach {\tool} is an API, which can be treated as a token sequence,
%Since the output of our API completion task is \cx{?} an API sequence,
we find utilizing beam search~\cite{sutskever2014sequence,wiseman2016sequence} can improve the performance. Beam search can return a list of the most likely output sequences, which can provide the developer with a few of the most likely API completion suggestions. 
%Specifically, it selects the $m$ tokens with the lowest heuristic cost at each time step, where $m$ denotes the beam width, by scanning through each step's API tokens one by one. It continues to select probable tokens for the following tokens after trimming any remaining branches until it comes across the end-of-sequence sign. Finally, for each description, our model may return $m$ candidate APIs.
%According to their average probabilities during the beam search operation, we rank the generated candidate APIs.
%Specifically, it scans through each step’s API tokens one by one and chooses the $k$ tokens with \cx{?}the lowest cost at each time step,
Specifically, it uses the lowest cost $k$ tokens of each time step by scanning the API tokens of each step one by one,
where $k$ denotes the beam width. After pruning any remaining branches, it continues to choose potential tokens for the subsequent tokens until it encounters the end-of-sequence sign. Finally, our model can return $k$ candidate APIs for each query. 
We rank the generated candidate APIs according to their average probabilities during the beam search procedure.

\subsection{Model Application Part}

In this part, given the query content and prompt information provided by developers, our  trained model can recommend a complete API.

% our trained model can recommend a complete API for developers based on the query content and prompt information provided by developers, including natural language descriptions of the required API and partial content of the API.

%% file: 4-ex_setup.tex
\section{Experimental Setup}
\label{sec:setup}

\subsection{Research Questions}

\textbf{{\RQone}}

\textbf{Motivation.} In this RQ, we want to evaluate the quality of APIs generated by {\tool} by comparing {\tool} with state-of-the-art baselines in terms of two types of automatic performance measures.

% choose five automatic measures of two types to evaluate the quality of API generated by {\tool} and our considered baselines to demonstrate the effectiveness of {\tool}.

\textbf{{\RQtwo}}

\textbf{Motivation.} In this RQ, we want to conduct ablation experiments by investigating the performance impact on {\tool} of different adversarial training methods for data augmentation (i.e., FGSM~\cite{goodfellow2015explaining}, FGM~\cite{miyato2017adversarial}, PGD~\cite{madry2017towards}, and our designed {\atpart}).

\textbf{{\RQthree}}

\textbf{Motivation.} 
In this RQ, we want to investigate the impact of different pre-trained models (i.e., PLBART~\cite{ahmad2021unified}, UniXcoder~\cite{guo2022unixcoder}, CodeBERT~\cite{feng2020codebert}, and 
our considered CodeT5~\cite{wang2021codet5}) on the performance of {\tool}.

\textbf{{\RQfour}}

\textbf{Motivation.} 
% In this RQ, we want to investigate the performance impact of different pre-trained models on {\tool}. Therefore, we consider the other three popular pre-trained models from the NLP field.
In this RQ, we want to explore the impact of prompts on the performance of {\tool}.
% , an NLP system that provides an API for generating text responses to user queries. 
Investigating the role of prompts is essential to understand how to optimize the model's performance,  and reduce the amount of training data.

\subsection{Experimental Subjects}

%To show the effectiveness of {\tool}, we selected 33k high-quality pairs from BIKER~\cite{huang2018api} based on \cx{?} three heuristic rules as our experimental subject.
To show the effectiveness of {\tool}, we select 33k high-quality query and API pairs from BIKER~\cite{huang2018api} as our experimental subject.
% Specifically, we reuse the Stack Overflow (SO) data from the official data dump of the SO method BIKER and design three rules by following previous suggestions from previous studies~\cite{bajaj2014mining,islam2019comprehensive} and the characteristics of this task. 
% These three heuristic rules are designed as follows.

% \begin{itemize}
%     \item \textbf{Rule 1:} A 1:n data pair is converted into multiple 1:1 data pairs.
%     \item \textbf{Rule 2:} Merge duplicate data to eliminate duplicates.
%     \item \textbf{Rule 3:} The length of APIs should not be smaller than 3 words.
% \end{itemize}

% After these three heuristic rules, we got a high-quality corpus for our empirical study.
The statistical information of the corpus is shown in 
 \tablename~\ref{tab:table2}.
In this table, we can find that the length of most queries and the corresponding APIs are mainly around 11 words and 4 words in the corpus after using the BPE algorithm. Moreover, 99.9\% of queries are less than 48 words, and 99.9\% of APIs are less than 16 words.
In our empirical study, we utilize a random sampling method to split the corpus into a training set, a validation set, and a test set in the ratio of 80\%:10\%:10\% by following previous studies~\cite{yang2021fine,liu2020retrieval}.

\begin{table}[htbp]
	\centering
\setlength{\tabcolsep}{3 mm}
% \vspace{-0.25cm}
\caption{Length statistics of our experimental subject}  
  % \vspace{-0.25cm}
  \resizebox{0.475\textwidth}{!}{
\begin{tabular}{cccccc}    
\toprule 
\multicolumn{6}{c}{Query length statistics}\\    
\midrule  Average  & Mode & Median &$<$16&$<$32&$<$48 \\
11.086 & 9 & 10 &84.2\% & 99.8\% &99.9\% \\ 
\midrule
\multicolumn{6}{c}{API length statistics}\\
\midrule  Average  & Mode & Median &$<$8&$<$12&$<$16 \\
4.713 & 4 & 4 & 94.9\% & 99.8\% &99.9\% \\ 
\bottomrule   
\end{tabular}} 
% \vspace{-0.45cm}
\label{tab:table2}
\end{table}

\subsection{Evaluation Measures}

% To compare the performance between {\tool} and baselines, we consider two performance measures (i,e., MRR~\cite{radev2002evaluating} and MAR~\cite{schutze2008introduction}). These performance measures have been widely used in previous API recommendation~\cite{huang2018api,rahman2016rack,wei2022clear} studies. 
% However, due to the short and clear characteristics of the API, these measures widely used for information retrieval~\cite{huang2018api,rahman2016rack,wei2022clear,lam2015combining,zanjani2015automatically} are difficult to accurately evaluate our API completion task by generation technology. Therefore, we consider using the EM metric~\cite{kenton2019bert,clark2020electra}, which measures whether the generated text by a model exactly matches the reference answer.
% The higher their scores, the better the performance of the model. 

To compare the performance of {\tool} and the baselines, we first consider two performance measures: Mean Reciprocal Rank (MRR)~\cite{radev2002evaluating} and Mean Average Precision (MAP)~\cite{schutze2008introduction}. These two measures have also been widely used in previous API recommendation studies~\cite{huang2018api,rahman2016rack,wei2022clear}.

\textbf{MRR.}
MRR~\cite{radev2002evaluating} measures the model's ability to rank multiple candidate answers by calculating the average of the reciprocal of the best rank for each query, which is defined as follows:
%Typically, a smaller reciprocal value indicates a higher rank, so a higher MRR indicates better ranking performance of the model. 
\begin{equation}
    MRR = \frac{1}{\left|Q\right|} \sum_{i=1}^{\left|Q\right|} \frac{1}{rank_i}
    \label{eq:MRR}
\end{equation}
where $Q$ represents the total number of questions in the evaluation set, and $rank_i$ represents the position of the first correct answer in the answer set predicted for the $i$-$th$ question.

\textbf{MAP.}
MAP~\cite{schutze2008introduction} measures the average precision score of the Precision-Recall curve, which represents the trade-off between precision and recall at different thresholds. Therefore, the MAP can be defined as follows:
%The MAP is calculated by ranking search results by relevance, determining a threshold for identifying relevant documents, computing the Precision-Recall curve, and calculating the area under the curve.
\begin{equation}
    MAP = \frac{1}{\left|Q\right|} \sum_{i=1}^{\left|Q\right|} AveP \left( {C_i},{A_i} \right)
    \label{eq:MAP}
\end{equation}
where $Q$ refers to the number of queries and $AveP(C_i,A_i)$ refers to the average accuracy.

% \cx{show how we compute MRR and MAR by {\tool}}

% % \gyf{As is shown in \eqref{eq:MRR}, $Q$ represents the total number of questions in the evaluation set, 
% and $rank_i$ represents the position of the first correct answer in the answer set predicted for the $i$th question. 
% If the answer set $C_i$ generated for the question $Q_i$ does not include the correct answer $A_i$, the value of $\frac{1}{rank_i}$ is set to 0.}

% where $Q$ is   $rank_i^j$
% The higher the position of the correct answer, the better. For example, the first predicted answer is the golden answer, which is equal to 1. The more inaccurate the value, the smaller it will be. Therefore, the larger the MRR value, the better

% \gyf{As is shown in \eqref{eq:MAP}, Where $Q$ refers to the number of queries, $AveP(C_i,A_i)$ refers to the average accuracy.}

% where $Q$ is   $rank_i^j$

Moreover, since we model the API recommendation problem as the automatic completion task, we further consider the Exact Match (EM) measure~\cite{kenton2019bert,clark2020electra}, which measures whether the API generated by the model matches the ground-truth API. 
Notice EM@k indicates the correct rate when comparing the top $n$ recommendations by using beam search.
Since it is useless to measure the text similarity in the API recommendation task, some popular evaluation measures (such as BLEU~\cite{papineni2002bleu} and ROUGE~\cite{rouge2004package}) used in the text generation tasks are not considered in our study.
% We illustrate the details of these performance measures as follows.

% \textbf{EM.}
% EM~\cite{kenton2019bert,clark2020electra} measures whether the generated text by a model exactly matches the reference answer. It also considers all words and punctuation as important information and only assigns high scores when the generated text matches the reference answer exactly. 
% EM@k indicates the correct rate of comparing the top $n$ predictions.
%The final EM@k value is the number of correctly predicted samples divided by the total number of samples.
%There is no doubt about the correctness of API in our API completion task, so we consider EM@k as the main evaluation metric.

For these two types of measures, the higher the score, the better performance of the corresponding approach.
To avoid result differences due to different implementations for performance metrics, we use the implementations provided by the nlg-eval library\footnote{https://github.com/Maluuba/nlg-eval} for these performance measures, which can mitigate the threat to the internal validity of our empirical study.

\subsection{Baselines}
% Since there are no previous studies on API completion, to show the competitiveness of our proposed method {\tool}, we compare our proposed method with state-of-the-art baselines about API recommendation methods\cite{huang2018api,rahman2016rack,wei2022clear} and pre-trained language models \cite{feng2020codebert,guo2022unixcoder,ahmad2021unified}.  

To demonstrate the competitiveness of our proposed approach {\tool}, we first compared {\tool} with state-of-the-art API recommendation baselines,
% , as there has been no previous research on API complementation.
% We first evaluated classical API recommendation methods\cite{huang2018api,rahman2016rack,wei2022clear}, including 
such as BIKER~\cite{huang2018api}, RACK~\cite{rahman2016rack}, and CLEAR~\cite{wei2022clear}. We introduce the characteristics of these baselines as follows.

\begin{itemize}

\item \noindent \textbf{BIKER.} BIKER \cite{huang2018api} uses word embedding technology to bridge the vocabulary gap and uses API to supplement the information to bridge the knowledge gap.

\item \noindent \textbf{RACK.} RACK \cite{rahman2016rack} uses keywords and API associations in stack overflow crowd-sourcing knowledge to recommend a list of related APIs for natural language queries.

\item \noindent \textbf{CLEAR.} CLEAR \cite{wei2022clear} uses a BERT-based model to embed whole sentences of queries and stack overflow posts, and uses contrastive learning to train the model.

\end {itemize}

Since we model the API recommendation problem as the automatic completion task, we also compare {\tool} with baselines based on the pre-trained language models, such as CodeBERT~\cite{feng2020codebert}, UniXcoder~\cite{guo2022unixcoder}, and PLBART~\cite{ahmad2021unified}.
% \cite{feng2020codebert,guo2022unixcoder,ahmad2021unified}
We introduce the details of these  pre-trained language models as follows.

\begin{itemize}

\item \noindent \textbf{CodeBERT.} CodeBERT~\cite{feng2020codebert} trains the model based on the neural architecture of Transformer and uses the mixed objective function.

\item \noindent \textbf{UniXcoder.} UniXcoder~\cite{guo2022unixcoder} is a unified cross-modal pre-trained model, which uses a masked attention matrix to control the model and uses cross-modal content to enhance code representation.

\item \noindent \textbf{PLBART.} PLBART~\cite{ahmad2021unified} is a text-to-text model, which learns program syntax, style, and logical flow that are crucial to program semantics.

\end {itemize}

%\syh{For baselines that do not share code \cite{huang2018api,rahman2016rack,wei2022clear}, we directly reuse the results of the original paper.}
For the first type of baseline~\cite{huang2018api,rahman2016rack,wei2022clear},  we implemented them according to the approach description, and the results of our implementations are very close to the results reported in these original studies.
For the second type of baseline~\cite{feng2020codebert,guo2022unixcoder,ahmad2021unified}, we directly use the script shared by these original studies. 
To ensure a fair comparison between {\tool} and these baselines, we optimized the hyper-parameters in these approaches.

\subsection{Implementation Details}

Our proposed approach and some baselines were implemented based on the PyTorch framework. 
We use the packages Transformers\footnote{\url{https://github.com/huggingface/transformers}} and HuggingFace \footnote{\url{https://huggingface.co}} to implement {\tool}.
To alleviate the overfitting issue, we employ an early stopping strategy that terminates training when the validation loss does not decrease for five consecutive iterations and saves the model's parameters with the best performance. 
The detailed hyper-parameter setting is shown in Table~\ref{tab:Hyper-parameters}. 
For the hyper-parameter of baselines, we adopt the value setting of original papers.
    
Our experiments were conducted on a computer with an Intel (R) Core(TM) i5-13600K and a GeForce RTX4090 GPU with 24 GB memory.

\begin{table}[htbp]
 \caption{Hyper-parameters setting of our proposed approach {\tool}}
 \begin{center}
\setlength{\tabcolsep}{6.4mm}
% \vspace{-5mm}
  \resizebox{0.475\textwidth}{!}{
\begin{tabular}{ccc}
\toprule
        \textbf{Component}               & \textbf{Hyper-parameter} & \textbf{Value} \\ \midrule
\multirow{2}{*}{{\atpart}} & $K$         & 4  \\
                       & $\alpha$         & 0.3    \\
                      \midrule
\multirow{5}{*}{CodeT5} & decoder\_layers       & 12     \\
                        & hidden\_size        & 768     \\
                       & max\_input\_length        & 48     \\
                       &max\_output\_length         & 16  \\
                       &beam\_search\_size        & 10  \\
  \bottomrule
\end{tabular}} 
% \vspace{-5mm}
 \end{center} 
 \label{tab:Hyper-parameters}
\end{table}

%% file: 5-ex_result.tex
\section{Experimental Results}
\label{sec:result}

\subsection{RQ1: Comparision with baselines}

% \textbf{Approach.}
In this RQ, we aim to investigate the competitiveness of our proposed approach {\tool}. Specifically,
% Since there are no previous studies on API completion, to show the competitiveness of our proposed method {\tool},
we compare our proposed approach with state-of-the-art API recommendation baselines~\cite{huang2018api,rahman2016rack,wei2022clear} and baselines based on the pre-trained language models~\cite{feng2020codebert,guo2022unixcoder,ahmad2021unified}. 

% \textbf{Results.}
We first compare {\tool} and API recommendation baselines in terms of two automatic evaluation measures (i.e., MRR and MAR).
The comparison results are shown in Table~\ref{tab:RQ11Result} and we emphasize the best result in bold.
According to the experimental results, 
we find our proposed approach {\tool} can at least outperform the API recommendation baselines (i.e., CLEAR)  by 13.20\% in terms of MRR and {\tool} can at least outperform the API recommendation baselines (i.e., BIKER)  by 16.31\% in terms of MAP.
% \cx{rewrite! summarize why our approach can outperform IR-based baselines}
% These results suggest that our proposed method is more effective and accurate in recommending missing API calls in code snippets.
% %分析，与传统信息检索相比，我们进步在什么地方，才导致性能好的。
% This result indicates that our proposed {\tool} method represents a significant improvement over traditional information retrieval methods for API recommendation. One key factor contributing to our method's superior performance is its ability to capture both contextual information and code syntax simultaneously, enabling it to more accurately and effectively recommend missing API calls in code snippets.
This result indicates that our proposed approach {\tool} can show a significant improvement over traditional IR-based API recommendation baselines. 
% Specifically, the information retrieval-based methods are only able to retrieve APIs that exist in the training set. However, in real-world scenarios, 
% there may be APIs present in the test set that were not encountered during the training process, resulting in the inability to retrieve relevant results.易被抨击，因为找不到例子
% Our proposed method {\tool}, effectively addresses this issue.
Since APIs obtained by the IR recommendation method may not be what we need, {\tool} can generate APIs that are more in line with the actual needs by using prefix prompts in the model.

We second compare {\tool} and generation baselines based on the pre-trained language models in terms of EM performance measure.
The comparison results are shown in Table~\ref{tab:RQ12Result}, and we also emphasize the
best result in bold. According to the experimental results, we find
{\tool} can at least outperform this type of baselines by 40.02\%, 32.70\%, 29.52\%, 23.85\%, and 24.50\% for EM@1, EM@2, EM@3, EM@4, and EM@5, respectively.
% \cx{rewrite! summarize why our approach can outperform baselines based on the pre-trained models.}
% These findings demonstrate the superior effectiveness and accuracy of {\tool} in generating missing API calls in code snippets. 
% Furthermore, our qualitative analysis confirms that {\tool} can provide more accurate and relevant API calls compared to the baseline method, highlighting its potential for improving the efficiency and accuracy of API generation in software development.
% %分析，与其他预训练模型相比，我们进步在什么地方，才导致性能好的。
This result shows the potential of {\tool} in improving the accuracy of API generation in software development.
Specifically, compared to previous baselines based on the  pre-trained language models, our proposed approach can help to improve semantic diversity through data augmentation, resulting in APIs with richer semantic information. 
Moreover, we can better align with user intent by utilizing prompt-based learning, resulting in APIs that are closer to their needs.

\begin{table}[]
 \caption{Comparison results between {\tool} and API recommendation baselines}
 \begin{center}
  % \vspace{-5mm}
 \setlength{\tabcolsep}{1mm}{
  \resizebox{0.475\textwidth}{!}{
\begin{tabular}{ccc|ccc}
  \toprule
\textbf{Approach Name} & \textbf{MRR} & \textbf{MAP} & \textbf{Approach Name} & \textbf{MRR} & \textbf{MAP} \\
  \midrule
% BIKER & 0.629 & 0.659 & CLEAR & 0.632 & 0.640  \\
% RACK  & 0.296 & 0.266 & \textbf{{\tool}} &\textbf{0.661} & \textbf{0.713}\\

RACK  & 0.278 & 0.249 & BIKER & 0.586 & 0.613  \\
CLEAR & 0.591 & 0.602 & \textbf{{\tool}} &\textbf{0.669} & \textbf{0.713}\\
  \bottomrule
\end{tabular}}
 }
 \end{center}
 % \vspace{-2mm}
 \label{tab:RQ11Result}
\end{table}

\begin{table}[]
 \caption{Comparison results between {\tool} and baselines based on the pre-trained language models}
 \begin{center}
  % \vspace{-5mm}
 \setlength{\tabcolsep}{1mm}{
  \resizebox{0.475\textwidth}{!}{
\begin{tabular}{c|ccccc}
  \toprule
\textbf{Approach Name} & \textbf{EM@1} & \textbf{EM@2} & \textbf{EM@3} & \textbf{EM@4} & \textbf{EM@5}\\
  \midrule

PLBART & 24.30 & 33.90 & 40.50 & 44.70 & 47.40 \\
UniXcoder & 33.00 & 45.60 & 52.00 & 55.70 & 57.70 \\
CodeBERT & 40.10 & 49.70 & 54.70 & 58.90 & 60.00 \\

  \midrule
\textbf{{\tool}} &\textbf{56.15} & \textbf{65.95} & \textbf{70.85} & \textbf{72.95} & \textbf{74.70}\\
  \bottomrule
\end{tabular}}
 }
 \end{center}
 % \vspace{-2mm}
 \label{tab:RQ12Result}
\end{table}

%总结，一两句话就行，文本中也就三四行
% In these two tables, we can find that our proposed method {\tool} can achieve the best performance in terms of all performance measures. 

% In conclusion, our proposed {\tool} method is highly competitive and outperforms the state-of-the-art baselines in both API recommendation and generation. These results demonstrate the effectiveness and potential of our method in practical applications in software development.

\begin{tcolorbox}[width=\linewidth, boxrule=0pt, top=1pt, bottom=1pt, left=1pt, right=1pt, colback=gray!20, colframe=gray!20]
\textbf{Summary to RQ1:}
{\tool} can outperform state-of-the-art API recommendation baselines and pre-trained language models in terms of MRR, MAP, and EM performance measures.
\end{tcolorbox}

\subsection{RQ2: Ablation study on adversarial training approaches}

% \begin{table}[]
%  \caption{Comparison results between CodeT5 and pre-trained language models}
%  \begin{center}
%   \vspace{-5mm}
%  \setlength{\tabcolsep}{1mm}{
%   \resizebox{0.475\textwidth}{!}{
% \begin{tabular}{c|ccccc}
%   \toprule
% \textbf{Method Name} & \textbf{EM@1} & \textbf{EM@2} & \textbf{EM@3} & \textbf{EM@4} & \textbf{EM@5}\\
%   \midrule
% PLBART & 22.35 & 28.70 & 32.75 & 35.60 & 38.25 \\
% UniXcoder & 38.90 & 49.30 & 54.80 & 59.10 & 61.55 \\
% CodeBERT & 47.15 & 58.25 & 64.35 & 67.30 & 69.50 \\
%   \midrule
% \textbf{{\tool}} &\textbf{56.70} & \textbf{66.40} & \textbf{70.35} & \textbf{73.35} & \textbf{75.05}\\
%   \bottomrule
% \end{tabular}}
%  }
%  \end{center}
%  \label{tab:RQ12Result}
% \end{table}
%消融实验
% \textbf{Approach.}
In this RQ, we aim to investigate the performance impact on {\tool} of different adversarial training methods (i.e., FGSM~\cite{goodfellow2015explaining}, FGM~\cite{miyato2017adversarial}, PGD~\cite{madry2017towards}, and our designed {\atpart}) and show the competitiveness of {\atpart}.
% In this RQ, we design some control approaches to verify the contributions of different adversarial training methods (i.e., FGSM~\cite{goodfellow2015explaining}, FGM~\cite{miyato2017adversarial}, PGD~\cite{madry2017towards}, and {\atpart}) for {\tool}.
Specifically,
we use $w/o \ AT$ to denote the control approach, which does not consider the adversarial training method for {\tool}.
In a similar way, we use FGSM (or FGM, PDG, and {\atpart}) to denote the control approach, which considers the corresponding adversarial training method for {\tool}.
To guarantee a fair comparison, {\tool} and other control approaches follow the same experimental setup.
We show the details of these adversarial training methods  as follows.

\begin{itemize}
    \item \textbf{FGSM.}
FGSM~\cite{goodfellow2015explaining} is an adversarial training method proposed by Goodfellow et al.~\cite{goodfellow2015explaining}. The input gradient is $g=\nabla _x L(f_\theta (x),y)$, where $\theta$ denotes the model parameter value, $x$ denotes the model input, $y$ denotes the label, and $L()$ denotes the loss function of the training model. The perturbation goes to the maximum of the loss function along the gradient direction, which is expressed as $\delta = \epsilon ·sign(g)$, where $sign()$ is the regularization method, and $\epsilon$ is the constraint that the perturbation is limited by infinite norm (i.e., $\left \| \delta \right \| _\infty <  \epsilon $).

\item \textbf{FGM.}
FGM~\cite{miyato2017adversarial} was proposed by Goodfellow et al.~ \cite{miyato2017adversarial}. Unlike FGSM, which takes the $sign$ function to regularize the gradient, FGM carries out $L2$ regularization on the gradient.
In addition, FGSM takes the same step in each direction, while FGM scales according to the specific gradient to get better adversarial examples. The perturbation is expressed as $\delta =\epsilon ·(g/\left \| g \right \|_2 )$, where $\epsilon$ is the constraint of perturbation ($L2$ norm of the distance between the original example and the adversarial example is always $\epsilon$).

\item \textbf{PDG.}
FGM calculates the perturbation directly through the  parameter $\epsilon$, which may not be optimal. Therefore, PGD ~\cite{madry2017towards} was improved and iterated several times to find the optimal perturbation.
The input gradient of each step is expressed as $g_t=\nabla _{x_t} L(f_\theta (x_t),y)$ and the perturbation of each step is expressed as $\delta_t =\epsilon ·(g_t/\left \| g_t \right \|_2 )$.
In the iteration, $\delta_t$ is gradually accumulated, and only the gradient calculated by the last $x_t+\delta_t$ is used when the parameters are finally updated.
\end{itemize}

In Table~\ref{tab:RQ2Result}, we show the ablation study results between {\atpart} and other classical adversarial training methods in terms of EM performance measure and mark the best result in bold.

\begin{table}[]
 \caption{Ablation study results between {\atpart} and other adversarial training methods}
 \begin{center}
  % \vspace{-5mm}
 \setlength{\tabcolsep}{1mm}{
  \resizebox{0.475\textwidth}{!}{
\begin{tabular}{c|ccccc}
  \toprule
\textbf{AT Method} & \textbf{EM@1} & \textbf{EM@2} & \textbf{EM@3} & \textbf{EM@4} & \textbf{EM@5}\\
  \midrule
w/o \ AT & 52.60 & 61.65 & 66.30 & 68.75 & 70.60 \\
with FGSM & 53.39 & 62.31 & 66.90 & 69.15 & 70.75 \\
with FGM & 54.85 & 64.00 & 67.65 & 70.55 & 72.15 \\
with PDG & 55.15 & 64.70 & 68.70 & 70.70 & 72.40 \\
  \midrule
\textbf{with {\atpart}} &\textbf{56.15} & \textbf{65.95} & \textbf{70.85} & \textbf{72.95} & \textbf{74.70}\\
  \bottomrule
\end{tabular}}
 }
 \end{center}
 \label{tab:RQ2Result}
\end{table}

%发现加不加对抗，分析一波
Firstly, we find that {\tool} with classical adversarial training (AT) outperformed {\tool} without AT in all cases.
Specifically, when using FGSM, FGM, and PGD methods, the performance of {\tool} with AT can be Significantly improved by at least 1.50\%, 4.28\%, and 4.85\% in terms of EM@1 measure.
This shows that the use of Adversarial Training Augmentation can effectively enhance the quality of the generated APIs in the API completion task.
% , respectively, compared to the approach without AT.%which metric
% Furthermore, compared to xxx, {\tool} can improve the performance by xx\% and xx\% for MRR and MAR, respectively.
% %分析，与不加对抗相比，我们进步在什么地方，才导致性能好的。
% This result indicates that incorporating AT into {\tool} can effectively enhance its ability to capture more informative features and improve its overall performance, as demonstrated by the significant increase in accuracy for both MRR and MAR.
%发现加不同的对抗，分析一波

Secondly, we find that {\tool} with our designed method {\atpart} can outperform the other classical AT methods.
Specifically, the performance of {\tool} with {\atpart} can be at least improved by 5.17\%, 2.37\%, and 1.81\% when compared to {\tool} with FGSM, FGM, and PGD.
% Specifically, compared to xxx, {\tool} can improve the performance by xx\% and xx\% for MRR and MAR, respectively, when using our proposed {\atpart} method. 
% This comparison result indicates that our designed method {\atpart} can effectively improve the ability of {\tool} to \cx{modify} capture more informative features and improve its overall performance. 
The comparison results show that {\atpart}  can improve the performance while ensuring stability by using both $L1$ and $L2$ regularization methods to optimize the adversarial loss. 
Therefore, our proposed new adversarial training method {\atpart} can effectively improve the overall performance of {\tool}.
% as demonstrated by the significant increase in accuracy for both MRR and MAR.
%分析，与其他对抗相比，我们进步在什么地方，才导致性能好的。
% Additionally, our proposed {\atpart} method is more effective than other AT methods in improving the performance of {\tool}.

% Overall, our study shows that adversarial training is effective in improving {\tool} performance, with {\atpart} being the most effective method. 
% This suggests that {\atpart} has great potential for enhancing {\tool} in practical applications. 
% %See Table~\ref{tab:RQ2Result} for specific accuracy improvements.
% For specific accuracy improvements, please refer to Table~\ref{tab:RQ2Result}.

\begin{tcolorbox}[width=\linewidth, boxrule=0pt, top=1pt, bottom=1pt, left=1pt, right=1pt, colback=gray!20, colframe=gray!20]
\textbf{Summary to RQ2:}
According to the results of our ablation experiment, we find using  our designed adversarial training method {\atpart} can help to achieve the best performance for {\tool}.
\end{tcolorbox}

\subsection{RQ3: Ablation study on pre-trained models}
% \textbf{Approach.}
% To showcase the effectiveness of CodeT5 in the API complementation task, we compared it with three pre-trained models selected in RQ1. To ensure a fair comparison, we maintained the same experimental setup for the fine-tuning phase of all four models.
In this RQ, we aim to demonstrate the effectiveness of using CodeT5 for {\tool}. 
To answer this RQ, we conducted a comparative analysis with the other three classical pre-trained models discussed in Section 4.4. 
To ensure a fair comparison, we used the same experimental setup in the model fine-tuning phase (details can be found in Section 3.2.2) for these pre-trained models. 
% This approach allowed us to directly evaluate and compare the performance of each model on the API complementation task.

% \textbf{Results.}
% Table~\ref{tab:RQ31Result} shows the results of our experiments on the effect of different pre-trained models. To highlight the best model for each metric, we have marked it in bold. Based on our findings, we observed that the CodeT5 model outperformed the other three pre-trained models. Specifically, CodeT5 significantly improved the performance of {\tool} by at least 20\%, 29\%, 9\%, 9\%, 9\% for EM@1, EM@2, EM@3, EM@4, and EM@5 compared to the three other models. These results suggest that CodeT5 is better suited for the task at hand than the other pre-trained models.
% Overall, our results suggest that CodeT5 is a highly effective pre-trained model for code generation, outperforming other state-of-the-art models such as PLBART, UniXcoder, and CodeBERT under various ablation conditions. This finding has important implications for the development of natural language processing tools for code generation, as it suggests that CodeT5 may be a promising approach for achieving high-quality code generation results in practical applications.
We use Table~\ref{tab:RQ31Result} to show the performance impact of different pre-trained models and highlight the best performance in bold.
The comparison results show that {\tool} with  CodeT5 can outperform the other three classical pre-trained models.
Specifically, compared to the other pre-trained models, CodeT5 can at least improve the performance by 19.09\%, 13.22\%, 10.10\%, 8.40\%, and 7.48\% in terms of  EM@1, EM@2, EM@3, EM@4, and EM@5, respectively. 
These results imply that CodeT5 is better suited for the task of API complementation than the other pre-trained models.

% Overall, our results demonstrate that CodeT5 is a highly effective pre-trained model for code generation, consistently outperforming state-of-the-art models such as PLBART, UniXcoder, and CodeBERT across different ablation conditions. This finding has significant implications for the development of natural language processing tools for code generation, as it suggests that CodeT5 may hold promise for achieving high-quality code generation results in practical applications.

\begin{table}[]
 \caption{Ablation study results between CodeT5 and other pre-trained language models}
 \begin{center}
  % \vspace{-5mm}
 \setlength{\tabcolsep}{1mm}{
  \resizebox{0.475\textwidth}{!}{
\begin{tabular}{c|ccccc}
  \toprule
\textbf{Approach Name} & \textbf{EM@1} & \textbf{EM@2} & \textbf{EM@3} & \textbf{EM@4} & \textbf{EM@5}\\
  \midrule

PLBART & 40.30 & 48.80 & 54.50 & 57.20 & 60.40 \\
UniXcoder & 38.90 & 49.30 & 54.80 & 59.10 & 61.55 \\
CodeBERT & 47.15 & 58.25 & 64.35 & 67.30 & 69.50 \\

  \midrule
\textbf{CodeT5} &\textbf{56.15} & \textbf{65.95} & \textbf{70.85} & \textbf{72.95} & \textbf{74.70}\\
  \bottomrule
\end{tabular}}
 }
 \end{center}
 \label{tab:RQ31Result}
\end{table}

\begin{tcolorbox}[width=\linewidth, boxrule=0pt, top=1pt, bottom=1pt, left=1pt, right=1pt, colback=gray!20, colframe=gray!20]
\textbf{Summary to RQ3: }
According to the results of our ablation experiment, we find using the recently proposed pre-trained model CodeT5 can help to achieve the best performance for {\tool}.

\end{tcolorbox}

\subsection{RQ4: Ablation study on using prompts}

%方法内提示长度的影响
% \textbf{Approach.}
% In this RQ, we investigate the performance impact of prompts with different lengths (such as 0 prefixs, 1 prefix, 2 prefixes, or random prefixs). 
% Specifically, $0 \ prefixs$ stands for no prompt and $1 \ prefix$ (or $2 \ prefixs$) stands for the first 1 prefix (or 2 prefixs) of the prompt. $random \ prefixs$ is the masking strategy used in this method, which generates different prompts with the last $n_{rand}$ prefix masked.
% To guarantee a fair comparison, {\tool} and other control approaches follow the same experimental setup.
In this RQ, we aim to investigate the impact of using prompts on the performance of {\tool}. Specifically, we compare the performance of models trained by {\tool} with and without prompts. 
% The prompts used in this study include no prompts and prompts with random prefixes. 
To ensure a fair comparison, we follow the same experimental setup for these two approaches. 
% By comparing the performance of the models trained with and without prompts, we can evaluate the impact of prompts on model performance.

% \textbf{Results.}
% Table~\ref{tab:RQ2Result} shows the automatic evaluation results concerning EM measure of our proposed adversarial training method {\atpart} and xxx. 
% We also mark the best one of each metric in bold.
% \gyf{Based on the results presented in Table~\ref{tab:RQ3Result}, the average accuracy of {\tool} with prompts was significantly higher than that of {\tool} without prompts . These findings suggest that using prompts can improve the performance of API completion models.
% % Moreover, we conducted further analysis to investigate the impact of using different types of prompts on {\tool}'s performance. The results showed that using prompts with random prefixes achieved the highest accuracy (XX\%). This is because random prefix prompts provide diverse and unpredictable context information, which helps the API completion model better understand the developers' intent and generate more accurate API suggestions.
% In contrast, when no prompts are used, the API completion model has no context information to rely on and may produce less accurate suggestions. Therefore, the results suggest that prompts can be a valuable tool in improving the performance of API completion models.}
We show the comparison results in Table~\ref{tab:RQ3Result}.
Specifically, {\tool} with prompts can improve the performance by 13.66\%, 11.21\%, 12.28\%, 12.06\%, 10.67\% in terms of EM@1, Em@2, EM@3, EM@4, EM@5 when compared with {\tool} without prompts.
These results show that incorporating prompts can provide valuable context information for the API completion model, which can eventually result in more accurate suggestions.

\begin{table}[]
 \caption{Comparison results between whether using prompt}
 \begin{center}
  \vspace{-5mm}
 \setlength{\tabcolsep}{1mm}{
  \resizebox{0.475\textwidth}{!}{
\begin{tabular}{c|ccccc}
  \toprule
\textbf{Approach Name} & \textbf{EM@1} & \textbf{EM@2} & \textbf{EM@3} & \textbf{EM@4} & \textbf{EM@5}\\
  \midrule
w/o prompt & 49.40 & 59.30 & 63.10 & 65.10 & 67.50 \\
% one prefix & 51.30 & 61.20 & 64.40 & 66.60 & 68.10 \\
% two prefixs & 56.00 & 66.40 & 70.80 & 72.90 & 74.00 \\  
   \midrule
\textbf{with prompt} &\textbf{56.15} & \textbf{65.95} & \textbf{70.85} & \textbf{72.95} & \textbf{74.70}\\
  \bottomrule
\end{tabular}}
 }
 \end{center}
 \label{tab:RQ3Result}
\end{table}

\begin{tcolorbox}[width=\linewidth, boxrule=0pt, top=1pt, bottom=1pt, left=1pt, right=1pt, colback=gray!20, colframe=gray!20]
\textbf{Summary to RQ4:}
According to the results of our ablation experiment, we find using prompts can help to achieve the best performance for {\tool}.
\end{tcolorbox}

%% file: 6-discussion.tex
\section{Discussion}
\label{sec:discussion}

\subsection{Hyper-parameter Analysis}

\begin{figure}[htbp]
	\centering
    % \vspace{-1mm}
	\includegraphics[width=0.9\linewidth]{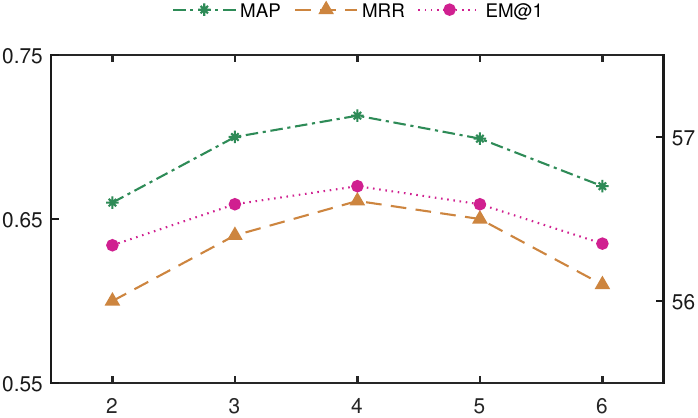}
	\caption{Sensitivity Analysis on the number of adversarial examples (i.e.,  the hyper-parameter $K$)}
    % \vspace{-5mm}
	\label{fig:kkk}
\end{figure}

%分析k
In this subsection, we perform a sensitivity analysis on the hyper-parameters of {\tool} to explore their optimal settings. 
Here we mainly focus on the hyper-parameter $K$, which is the number of adversarial examples. 
%The results of the sensitivity analysis are shown in \figurename~\ref{fig:kkk}, where all hyper-parameters except the hyper-parameter of the current analysis are set to the optimal setting. 
\figurename~\ref{fig:kkk} shows the results of the sensitivity analysis.
Except for the hyper-parameter of the current analysis, the values of all hyper-parameters are set to our optimal settings.
In this figure,
the left vertical axis is used for measuring MAP and MRR, and the right vertical axis is used for measuring EM@1.
Based on the sensitivity analysis results, we can find the  optimal value for the hyper-parameter $K$  is 4 for our investigated API completion task.

% \figurename~\ref{fig:kkk} shows that when the number of adversarial examples $k$ is 4, the scores of the three evaluation indicators shown are maximized.
% When the number of generated adversarial examples ranges from 2 to 4, the scores of these evaluation metrics first show a downward trend, followed by an upward trend.
% When the number of generated adversarial examples is greater than 4, the scores of these evaluation metrics show a flat decreasing trend.
% We set the number of generated adversarial examples to 4 to balance the performance and cost of the model in our experiments.

%图还没替换 

\subsection{Prefix Length Influence of {\tool}}

In this subsection, we want to analyze how the lengths of the prefixes  influence the performance of {\tool}. We conduct a sensitivity study on the performance impact of prompts with different prefix lengths (i.e., 0 prefix, 1 prefix, 2 prefixes). 
Specifically, 0 prefix stands for no prompt. 1 prefix (or 2 prefixes) stands for only providing the first one prefix (or the first two prefixes). The remaining parts follow the same experimental setup to guarantee a fair comparison. 

The final results are shown in Table~\ref{tab:Discussion6.2}.
The average accuracy of {\tool} with 2 prefixes was 74.00\%, which is significantly higher than that of {\tool} without prompts (67.50\%) and with 1 prefix (68.10\%) in terms of EM@5. 
Therefore, increasing the prefix length (i.e., providing more context information) can lead to better API completion suggestions.

\begin{table}[]
 \caption{The performance of {\tool} with different prefix lengths}
 \begin{center}
  % \vspace{-5mm}
 \setlength{\tabcolsep}{1mm}{
  \resizebox{0.475\textwidth}{!}{
\begin{tabular}{c|ccccc}
  \toprule
\textbf{Prefix Length} & \textbf{EM@1} & \textbf{EM@2} & \textbf{EM@3} & \textbf{EM@4} & \textbf{EM@5}\\
  \midrule
0 prefix & 49.40 & 59.30 & 63.10 & 65.10 & 67.50 \\
1 prefix & 51.30 & 61.20 & 64.40 & 66.60 & 68.10 \\
2 prefixes & \textbf{56.00} & \textbf{66.40} & \textbf{70.80} & \textbf{72.90} & \textbf{74.00} \\  
%   \midrule
%\textbf{with prompt} &\textbf{56.15} & \textbf{65.95} & \textbf{70.85} & \textbf{72.95} & \textbf{74.70}\\
  \bottomrule
\end{tabular}}
 }
 \end{center}
 \label{tab:Discussion6.2}
\end{table}

\subsection{Performance Improvement by API Correctness Check}
% 这个API check感觉确实没太大意义
%In this subsection, the API check module, {\APIcheck}, is designed to perform syntax checking on the 5 complete APIs generated by our proposed method {\tool} to ensure correctness. 
% 这个地方的五个生成值，指的是对于每一个query，模型返回了五个生成值，对他们进行check，原文稍微有点表述不清
In this subsection, we aim to further improve the performance of {\tool} by considering API correctness check.
Specifically, these APIs are submitted to an external API library to determine whether they have the correct syntax and semantics. The external API library contains all available APIs, which serve as a reference for syntax and semantic constraints. 
If the generated API cannot be found in the library, it is considered incorrect. 
For all APIs generated by beam search, we check them in sequence until we find five APIs, which exist in the external API library.

% 已将序列删掉

\tablename~\ref{tab:disAPIcheck} shows the final comparison  results.
In this table, we also concern EM measure and mark the best one of each measure in bold.
The experimental results show that by using {\APIcheck}, {\tool} can further improve the model performance. 
Specifically {\tool} can improve the performance by 5.79\%, 2.88\%, 0.99\%, 1.64\%, and 2.07\% in terms of EM@1, EM@3, EM@5, MRR, and MAP, respectively.
% This result indicates {\APIcheck} can help to self-check and filter the illegal API can help to further improve the performance of {\tool} by checking external API libraries.

\begin{table}[]
 \caption{Comparison results of whether to use the module {\APIcheck} or not}
 \begin{center}
  % \vspace{-5mm}
 \setlength{\tabcolsep}{1mm}{
  \resizebox{0.475\textwidth}{!}{
\begin{tabular}{c|ccccc}
  \toprule
\textbf{Approach Name} & \textbf{EM@1} & \textbf{EM@2} & \textbf{EM@3} & \textbf{EM@4} & \textbf{EM@5}\\
  \midrule
w/o {\APIcheck} & 56.15 & 65.95 & 70.85 & 72.95 & 74.70 \\
\midrule
\textbf{with {\APIcheck}}  &\textbf{59.40} & \textbf{67.85} & \textbf{71.55} & \textbf{74.15} & \textbf{76.25}\\
  \bottomrule
\end{tabular}}
 }
 \end{center}
 % \vspace{-5mm}
 \label{tab:disAPIcheck}
\end{table}

% 实验结果未完全出来

\subsection{Limitations of {\tool}}

After automatic evaluation, we can find that {\tool} can achieve better performance than baselines. 
However, we also notice that {\tool} may generate low-quality APIs when compared to ground truth. 
In this subsection, we identify three challenge types  and analyze them in details.

% \gyf{One type of challenge is when the user is unfamiliar with the target API or has spelling errors, which can result in error prefixes. This indicates that the API generated by the prefix does not meet the user's needs.
% For example, the given query is' How do I modify the setBounds method for JComponents? '. 
% The user should have entered the prefix information as' java. awt', but the user mistakenly entered the prefix as' javas. awt '. The API generated by {\tool} is' javas. awt. component. setbounds'. 
% However, its corresponding ground truth should be' java. awt. component. setbounds'.
% A possible solution is to use a fuzzy matching algorithm to match the prefix entered by the user. 
% When the user enters an incorrect API prefix, the system can use a fuzzy matching algorithm to match the closest prefix to generate the correct API.}

The first challenge type is that developers may not accurately describe their queries, which may result in poor-quality generated APIs. 
For example, developers want to find what is the usage of Interrupts in Java,
The ground truth is "java. lang. Thread. stop", but the developers incorrectly describe it as "how Java is partitioned", resulting in the generation of incorrect APIs. 
One possible solution is to use query reformulation methods~\cite{cao2021automated}  to alleviate this issue.

% \gyf{The second challenge type is users may not accurately describe the query they input, resulting in the generated API not meeting the user's usage requirements.}
% % when longer APIs are present in the dataset, the random mask method we use may result in relatively short input prompts. 
% % As a consequence of insufficient prompt information, the generated API may not meet the user's requirements.
% \gyf{For example, } 

% \ysy{The second challenge type is that users can not provide accurate prompts or are even unable to provide prompts. 
% In this scenario, {\tool} can not complete the incomplete API. 
% For example, given the query "Java Thread suspend does not work" and the prompt "java. lang". 
% Originally, {\tool} can generate ground truth "Thread. sleep". 
% However, if users can not provide the prompt "java. lang", {\tool} would generate the API "java.lang.thread.suspend". 
% A possible solution is a system to provide an optional list of API prefixes, from which users can choose an API prefix that matches their needs and memory.}

The second challenge type arises when the required API consists of a large number of tokens after using the BPE algorithm. In such cases, even if the developer correctly enters the prefix, there is still a possibility of generating the wrong API. 
For example, the developer wants to create HTML documents from HTML strings in Java, and the corresponding ground truth should be "javax.swing.text.html.HTMLDocument.setOuterHTML".
Even though the prefix given by the developer is "Javax.swing", the  API generated by {\tool} is "javax.swing.jeditorpane.setpage", which is different from the ground truth.
% One possible solution is to set up a detection system that prompts users to increase the number of prompts when encountering longer APIs. 

The third challenge type is related to the trained model. This shows there still exists performance improvement room for the current model. 
Therefore, further augmenting the training data and improving the diversity of the gathered queries and APIs may alleviate this issue.

% \gyf{The third possible issue is that we
% One possible solution is to perform usability checks on the generated APIs to ensure that all APIs output to users are available}s 

\subsection{Threats to Validity}

\textbf{Internal threats.}
The first internal threat is the potential errors during the implementation of {\tool} and the baselines. 
To mitigate this threat, we use mature frameworks (such as PyTorch and Transformer) to ensure the correctness of our code implementation. 
For the baseline, we use scripts shared in previous work to alleviate this threat.
The second threat is related to our hyper-parameter settings. 
It may take a huge computational csot to find the optimal hyper-parameter settings for {\tool}. However, based on our current hyper-parameter settings, {\tool} can still achieve better performance than baselines.
% 我们面临的第一个内部威胁是{\tool}和基线实现中可能存在的潜在错误。为了减轻这种威胁，我们使用了成熟的框架，如PyTorch和Transformer，以确保代码实现的正确性。对于基线，我们使用之前的工作中共享的脚本。第二个内部威胁与各种超参数设置有关。在寻找{\tool}的最佳超参数设置时可能需要花费很长时间。但是，根据我们目前的超参数设置，{\tool}仍然可以取得很好的性能，并且可以通过超参数优化进一步提高性能。

\textbf{External threats.}
The first external threat is related to the quality of our constructed corpus.
To alleviate this threat, we use the experimental subject shared by Huang et al.~\cite{huang2018api}, which can guarantee the generalization of our empirical findings. 
The second threat is we only use the dataset shared by Huang et al.~\cite{huang2018api} to demonstrate the effectiveness of {\tool}, which only supports Java programming language. Therefore, the performance of {\tool} may differ when performing API recommendations for other programming languages, especially some low-resource programming languages (such as Ruby, and Go).

\textbf{Construct threats.}
The main construct threat is related to our used performance measures.
To evaluate the effectiveness of {\tool}, we consider IR-based measures used by previous API recommendation studies~\cite{huang2018api,rahman2016rack,wei2022clear}.
Moreover, to evaluate our API completion method more accurately, we also consider EM metric~\cite{kenton2019bert,clark2020electra}, which can measure whether the generated API can completely match the ground-truth API.

% \textbf{Conclusion threats.}
% The primary threat to the conclusion of our human study is related to the potential bias introduced by the participants. To mitigate this risk, we first invited graduate students who are familiar with Stack Overflow to participate in our study. Secondly, we provided a tutorial to ensure that all graduate students understood our protocol. Lastly, we used the Fleiss Kappa measure to assess the consistency among all graduate students.
% 我们的人类研究的主要结论威胁与参与者可能引入的偏差有关。为了缓解这种风险，我们首先邀请熟悉Stack Overflow的研究生参与我们的研究。其次，我们提供了教程以确保所有研究生都理解我们的协议。最后，我们使用了Fleiss Kappa度量来评估所有研究生之间的一致性。

%% file: 7-conclusion.tex
\section{Conclusion}
\label{sec:conclusion}
In this study, we provide a new direction for API recommendation by modeling this problem as the automatic completion task.
In the completion task, the developer can input the query and the API prefixes (i.e., the prompt) they know. 
Then we propose a novel API completion approach {\tool} with prompt learning and adversarial training-based data augmentation.
Specifically, {\tool} trains the API completion model through API prompts and adversarial examples generated by  our designed method {\atpart} in the embedding layer.
In our experimental studies, we evaluate the effectiveness of {\tool} in terms of two types of performance measures. Final results show that {\tool} can outperform state-of-the-art API recommendation baselines and baselines
based on the pre-trained language models.
Finally, we also verify the rationality of the component setting in {\tool}, such as our designed adversarial training method, the used pre-trained model, and prompt learning.

In the future, we first want to apply our proposed approach to other programming languages to verify the effectiveness of {\tool}. We second want to improve the performance of {\tool} by considering more advanced data augmentation techniques and large language models (such as Codex or ChatGPT). Finally, we want to automatically generate parameters that are needed for invoking APIs, which can further improve the practicability of {\tool}.